\documentclass[preprint2]{aastex631}

\shorttitle{370 New Eclipsing Binary Candidates}
\shortauthors{Howard et al.}

\begin{document}

\title{370 New Eclipsing Binary Candidates from TESS Sectors 1-26}

\author[0000-0002-0716-947X]{Erin L. Howard}
\affiliation{Physics \& Astronomy Department, Western Washington University, 516 High Street, Bellingham, WA 98225, USA}

\author[0000-0002-0637-835X]{James R. A. Davenport}
\affiliation{Astronomy Department, University of Washington, Box 951580, Seattle, WA 98195, USA}

\author[0000-0001-6914-7797]{Kevin R. Covey}
\affiliation{Physics \& Astronomy Department, Western Washington University, 516 High Street, Bellingham, WA 98225, USA}

\begin{abstract}

We present 370 candidate eclipsing binaries (EBs), identified from ${\sim}$510,000 short cadence TESS light curves. Our statistical criteria identify 5,105 light curves with features consistent with eclipses (${\sim}$1\% of the initial sample). After visual confirmation of the light curves, we have a final sample of 2,288 EB candidates. Among these, we find 370 sources that were not included in the catalog recently published by \citet{prsa2022}.  We publish our full sample of 370 new EB candidate, and statistical features used for their identification, reported per observation sector. 
\end{abstract}

\section{Training Set} \label{sec:training}

To provide a training set for our selection criteria, we manually classified ${\sim}$2,500 randomly selected TESS light curves \citep{tess} to identify candidate EBs. Supplemented with additional EBs flagged from Kepler \citep{kirk2016}, our training set includes 64 EBs and 2,438 non-EBs.

\section{Light Curve Processing} \label{sec:processing}

We searched 2-minute cadence TESS light curves (LCs) from Sectors 1-26 archived at the Mikulski Archive for Space Telescopes. Data were downloaded in September 2020, and the analysis used PDSCAP fluxes. Data with a quality flag other than zero were omitted. To eliminate background subtraction errors, ${\sim}$1.5 days of data on either side of data gaps were removed, including the start and end of each sector's observation.

\section{Iterative Search Process} \label{sec:overview}

We performed an iterative search process, where in each of three iterations we:

\begin{enumerate}
    \item Search each LC for its deepest candidate eclipse signature;
    \item Use multiple algorithms to search for periodicity in the LC
    \item Apply statistical thresholds to identify confident EBs or non-EB classifications per sector;
    \item For ambiguous cases, subtract a smoothed copy of the LC to increase the contrast of sharp eclipse features, and iterate again with adjusted thresholds.
\end{enumerate}

\subsection{Eclipse Search} \label{sec:eclsearch}

To isolate potential eclipses, we calculated the  $Z$ score of every data point in the LC.
This identifies the significance of each deviation from the LC mean, calculated by normalizing the difference of each data point ($x$) from the mean ($\mu$), after normalizing by the LC's standard deviation ($\sigma$).

\begin{equation}
    Z = \frac{x-\mu}{\sigma}
\end{equation}

To identify the most significant potential eclipse signature in each LC, we took the datapoint with the lowest Z score and averaged it with two datapoints on either side for a total of five data points, a value we denote as $Z_5$. Any LC that did not have five data points was automatically rejected. 

\subsection{Period Search} \label{sec:persearch}

We used standard python libraries to measure the dominant period and maximum power as identified by the Lomb-Scargle (LS), Auto-correlation Function (ACF), and Box-Least Squares (BLS) methods.


\subsection{Subtractive Smoothing} \label{sec:smooth}

Each sector LC that could not be classified as confidently containing or not containing an eclipse was then processed to remove non-eclipse variations and enhance the sensitivity of the eclipse signature search.  The LC was normalized by the median. We then subtracted out a smoothed LC, created by a rolling-median with a window of 1/5th the ACF period, or 128 data points if no periodicity was recovered. This window was tuned based on analysis of the TESS LCs for the 64 Kepler EBs in our training set. 

The subtraction smoothing was intended to filter out variations over longer timescales, leaving only rapid/sharp eclipse-like features. This smoothing process was carried out through each iteration that we describe below.

\subsection{Criteria for Candidate EBs} \label{sec:criteria}

The criteria used to identify confident EB classifications in each stage of our iterative process were developed from statistics computed for the training set of ~2,500 sector LCs described in section \ref{sec:training}. 
The decision tree of our iterative classification process was:

{\bf Pre-classification:} all LCs with $Z_5 < 3$ were removed from consideration and labeled as non-EBs.

{\bf First Pass:}
\begin{itemize}
    \item Confident EBs:
    \begin{itemize}
        \item BLS max power $> 1,500$ OR $Z_5 > 10$
    \end{itemize}
    \item Confident non-EBs:
    \begin{itemize}
        \item ACF max power $< 0.05$ \& $Z_5 < 4$, OR
        \item BLS max power $< 100$ \& $Z_5 < 7$, OR
        \item BLS max power $< 60$.
    \end{itemize}
\end{itemize}

{\bf Second Pass:}
\begin{itemize}
    \item Confident EBs: 
    \begin{itemize}
        \item BLS max power $> 850$ OR $Z_5 > 8$
    \end{itemize}
    \item Confident non-EBs:
    \begin{itemize}
        \item ACF max power $< 0.05$ \& $Z_5 < 4$, OR
        \item BLS max power $< 60$.
    \end{itemize}
\end{itemize}

{\bf Third Pass:}
\begin{itemize}
    \item Confident EBs: 
    \begin{itemize}
        \item BLS max power $> 200$ OR $Z_5 > 6$
    \end{itemize}
    \item Confident non-EBs
    \begin{itemize}
        \item ACF max power $< 0.05$ \& $Z_5 < 4$, OR
        \item BLS max power $< 60$.
    \end{itemize}
\end{itemize}

\subsection{Identified EBs} \label{sec:ebs}

Applying these criteria to the full set of 507,776 TESS sector 1-26 LCs eliminates 98\% as likely non-EBs, and flags 5,105 candidate EB systems for visual inspection.  
After identifying and relabeling false positives, false negatives, and undetermined cases, 2,287 objects were identified as confident EBs by the combination of our statistical pre-classifier and visual examination.

Matching this list against the catalog assembled by \citet{prsa2022}, we find 370 sources that they do not flag as EBs: the properties of these 370 EB candidates are listed in Table \ref{tbl1} below.  This table includes the number of times the curve was smoothed before reaching classification, the BLS and ACF periodicity found and their respective powers, an eclipse depth significance metric ($Z_5$), classifier decision, and manual classification.

\section{Conclusion} \label{sec:conclusion}
We present 370 new EB candidates from the TESS 2-minute light curves in Sectors 1-26. Our iterative approach does not require multiple eclipses to be detected within the TESS observing windows, so we have many single-eclipse candidates in our sample. While we use periodic estimators in our decision tree, which are sensitive to eclipse-like morphologies (e.g. the Boxed-Least-Squares), we do not necessarily recover the actual orbital period for any given system. Many of the strongest periods recovered and presented in Table 1 may be an alias of the true period.

Our search for eclipsing systems is complementary to the larger effort to generate a complete catalog of EBs from TESS \citep{prsa2022}. We have not vetted pixel-level data from TESS for the EB candidates in Table 1, and eclipses recovered in some systems may be contamination from nearby EBs. As the TESS mission continues to gather observations, however, we expect many of these candidates to have subsequent eclipses measured, and their status as true EBs confirmed.

\begin{acknowledgments}

We thank Andrej Pr{\v{s}}a for helpful discussions of our EB catalog.

ELH was supported by the Distributed Research Experiences for Undergraduates (DREU) program, a joint project of the CRA Committee on the Status of Women in Computing Research (CRA-W) and the Coalition to Diversify Computing (CDC), which is funded in part by the NSF Broadening Participation in Computing program (NSF592BPC-A \#1246649).

JRAD acknowledges support from the DiRAC Institute at the University of Washington, supported through generous gifts from the Charles and Lisa Simonyi Fund for Arts and Sciences, and the Washington Research Foundation. JRAD acknowledges support from the Heising-Simons Foundation, and Research Corporation for Science Advancement for hosting the 2019 Scialog meeting. 

This project used data collected with the TESS mission, obtained from the MAST data archive at the Space Telescope Science Institute (STScI). Funding for the TESS mission is provided by the NASA Explorer Program. STScI is operated by the Association of Universities for Research in Astronomy, Inc., under NASA contract NAS 5–26555.
\end{acknowledgments}

\software{
Python, IPython \citep{ipython}, 
NumPy \citep{numpy}, 
Matplotlib \citep{matplotlib}, 
SciPy \citep{scipy}, 
Pandas \citep{pandas}, 
Astropy \citep{astropy}
}

\begin{deluxetable*}{cccccccccc}
\tabletypesize{\footnotesize}
\tablecolumns{10}
\tablewidth{0pt}
\tablecaption{ \label{tbl1} }
\tablehead{
\colhead{Obj-ID}&\colhead{Sector} &\colhead{Smooth} & \colhead{BLS Max Power} & \colhead{BLS Per} & \colhead{ACF Max Power} & \colhead{ACF Per} & \colhead{$Z_5$} & \colhead{Class}  & \colhead{Manual}
}
\startdata
TIC 1536984 &   008 &   1  &    111.4535901 &  17.21034217 & 0.014287451 &  8.605171086 & 15 & EB&        EB          \\    
TIC 1695417 &   010 &   2  &    78.72930678 &  11.38450362 & 0.097307025 &  5.692251811 & 8  & EB&        EB          \\  
TIC 2013258 &   019 &   1  &    71.50674986 &  11.58446877 & 0.066572325 &  5.792234385 & 11 & EB&        EB          \\
TIC 4207747 &   006 &   1  &    2575.737558 &  14.92155815 & 0.067189255 &  7.460779077 & 16 & EB&        EB          \\    
TIC 4735737 &   022 &   1  &    712.860559  &  13.18314207 & 0.012599779 &  6.591571035 & 18 & EB&        EB          \\    
TIC 7720507 &   025 &   1  &    5935.405967 &  1.57154827  & 0.297338213 &  1.57154827  & 4  & EB&        EB          \\    
TIC 7720507 &   026 &   1  &    7543.012942 &  1.577082358 & 0.379232628 &  1.577082358 & 4  & EB&        EB          \\    
TIC 8444713 &   024 &   2  &    222.420267  &  9.40838592  & 0.230382421 &  9.40838592  & 8  & EB&        EB          \\    
TIC 8444713 &   025 &   3  &    201.0417848 &  9.43870895  & 0.27075056  &  9.43870895  & 6  & EB&        EB          \\    
TIC 9146275 &   002 &   1  &    2664.920962 &  14.76084518 & 0.010876056 &  7.380422591 & 14 & EB&        EB          \\    
\enddata
\tablecomments{Per-Sector details for our 370 EB candidates. ACF and BLS periods are in days. A preview of the table is shown here for reference.}
\end{deluxetable*}

\bibliographystyle{aasjournal}

\end{document}